# Muon Tracing and Image Reconstruction Algorithms for Cosmic Ray Muon Computed Tomography

Zhengzhi Liu, Stylianos Chatzidakis, John M. Scaglione, Can Liao, Haori Yang, and Jason P. Hayward

*Abstract*—Cosmic ray muon-computed tomography (μCT) is a new imaging modality with unique characteristics that could be particularly important for diverse applications including nuclear nonproliferation, spent nuclear fuel monitoring, cargo scanning, and volcano imaging. The strong scattering dependence of muons on atomic number Z in combination with high penetration range could offer a significant advantage over existing techniques when dense, shielded containers must be imaged. However, μCT reconstruction using conventional filtered back-projection is limited due to the overly simple assumptions that do not take into account the effects of multiple Coulomb scattering prompting the need for more sophisticated approaches to be developed.

In this paper, we argue that the use of improved muon tracing and scattering angle projection algorithms as well as use of an algebraic reconstruction technique should produce muon tomographic images with improved quality – or require fewer muons to produce the same image quality – compared to the case where conventional methods are used. We report on the development and assessment of three novel muon tracing methods and two new scattering angle projection methods for μCT. Simulated dry storage casks with single and partial missing fuel assemblies were used as numerical examples to assess and compare the proposed methods. The simulated images showed an expected improvement in image quality when compared with more conventional techniques, even without muon momentum information, which should lead to improved detection capability, even for partial defects.

## I. Introduction

Cosmic radiation, originating mainly from outside our solar system, constantly bombards the upper layers of the atmosphere and creates extensive showers of secondary particles, including muons, that eventually reach sea-level [1]. Cosmic ray muons are charged particles, approximately 200 times heavier than the electron, generated naturally in the atmosphere. They rain down upon the earth at an approximate rate of 10,000 particles $m^{-2}$ $min^{-1}$ at sea level and energies higher than 1 GeV [2]. Using cosmic ray muons for imaging applications presents several potential advantages since dense well shielded materials can be imaged. Conventional methods for examining the interior of materials e.g., x-rays, are limited by the fact that they cannot penetrate very dense well-shielded objects while more sophisticated techniques such as the penetrating neutrons or the recently developed proton radiography necessitate the use of an expensive accelerator. Further, utilization of muons requires no radiological sources and their high energy (>1 GeV) and penetration range (>1 m in rock), makes use of shielding for smuggling special nuclear material practically ineffective.

With availability of detectors that can measure the positions and directions of individual muons before and after traversing an object under investigation cosmic ray muons have been applied to probe thick structures [3], underground tunnels [4], nuclear fuel debris location [5], cargo containers [6] and even used to differentiate material [7-9]. Another application of interest is imaging spent nuclear fuel stored in shielded dry casks [10-16]. Monitoring nuclear waste and having the ability to understand whether there have been changes to the spent nuclear fuel geometry after transportation is important for waste management system planning efforts. Recent efforts have shown that μCT has the potential to reconstruct the contents of commercial vertical and horizontal dry casks storing fuel assemblies and even identify missing fuel assemblies [17-19].

Challenges in μCT include low muon flux, difficulty in muon momentum measurement, and the tendency of muons to scatter in the target and thus blur the image. No direct information about the muon path traversing the medium under interrogation is available and some type of extrapolation or heuristic assumption is required for muon imaging. Early image reconstruction of simple test objects was initially performed using a geometry-based reconstruction algorithm, known as the Point-of-Closest-Approach (PoCA) [20] inspired by earlier work on nuclear scattering radiography [21]. Statistical

This work was supported by DOE Office of Nuclear Energy's Nuclear Energy University Programs under contract DE-NE0008292. This research was also partially sponsored by the Laboratory Directed Research and Development Program of Oak Ridge National Laboratory, managed by UT-Battelle, LLC, for the US Department of Energy.

Zhengzhi Liu and Jason P. Hayward are with the University of Tennessee, Knoxville, TN 37996 USA (email: zliu36@vols.utk.edu; jhayward@utk.edu).

Stylianos Chatzidakis and John M. Scaglione are with the Oak Ridge National Laboratory, Oak Ridge, TN 37831 USA (email: chatzidakiss@ornl.gov; scaglionejm@ornl.gov).

Can Liao and Haori Yang are with Oregon State University, Corvallis, OR 97331 USA (email: liaoc@onid.oregonstate.edu; haori.yang@oregonstate.edu).

reconstruction techniques based on information obtained from muon scattering and displacement improved image reconstruction at the expense of computation and memory usage. These early techniques did not make use of CT. Recent µCT reconstruction efforts relied on the use of a straight-line path approximation defined by the incoming muon trajectory [17, 18]. This assumption does not take into account the effects of multiple Coulomb scattering (MCS). Overall, conventional muon tomographic algorithms struggle when using unaltered muon trajectories along with simple scattering angle projection method and the obtained images are of poor spatial resolution. For robust µCT, efficient and flexible algorithms are needed to model the MCS process [22, 23] and accurately estimate the trajectory of a muon as it traverses an object [24].

In the present work, three different muon tracing methods were investigated along with two different scattering angle projection algorithms for µCT reconstruction. For a moderately difficult real world example, the algorithms were applied to understand the expected effects on the quality of imaging a dry nuclear fuel storage cask. A simulated VSC-24 dry storage cask [25] with a fuel assembly or partial assemblies missing was used as a numerical model to capture the main characteristics of the proposed algorithms. This paper discusses how the simulated image quality may be expected to impact the detection of a missing fuel assembly. The validation of the simulation workspace with actual experimental data is also described.

## II. COMPARISON BETWEEN X-RAY AND MUON CT

In conventional transmission-based medical computed tomography [26], x-rays are generated by an X-ray tube or linear accelerator and then collimated to form a quasi-parallel beam before irradiating a patient or an object. The flux is manually controlled and can vary depending on the application, e.g., several million photons per $cm^2$, and the photon trajectory is straight. The projection information is the transmission rate of x-rays, which provides integral information of the material crossed by the x-ray beam. The incident beam often has significant probability of experiencing Compton scattering within an object. Scattered x-rays either are not registered by detectors or are registered by bins other than the bins hit by the uncollided x-rays, causing noise in the signal.

Contrary to x-rays, cosmic ray muons are naturally generated from the decay of pions, which are the products of interactions between primary cosmic rays and upper atmospheric atoms. The result is an uncollimated flux of particles at a low flux rate of approximately 1 muon/$cm^2$/minute at sea level. In addition, muon flux depends strongly on zenith angle and altitude. As charged particles, when muons pass through matter, they lose energy via ionization and are deflected from their incident direction via MCS from nuclei [1]. Since the energy spectrum of muons is continuous, and the average range is sufficient to allow the majority of muons to pass through most objects, both differential attenuation and scattering could be used to provide signals and generate tomographic images of the stored contents. It is worth noting that the variance of the scattering angle is more sensitive to atomic number Z than attenuation [27].

In both x-ray and muon CT, filtered back-projection (FBP) and algebraic reconstruction techniques (ART) can be used to reconstruct objects under investigation. A comparison of x-ray CT and muon CT is shown in Table I. The non-straight muon path and the use of scattering angles instead of transmission necessitates the development of a new imaging framework that includes ray tracing and projection techniques that can be coupled with FBP or ART.

TABLE I
COMPARISON BETWEEN X-RAY AND MUON CT

|  | X-ray CT | Muon CT |
|---|---|---|
| Source | X-ray tube/Linac | Natural cosmic ray |
| Flux rate | Controllable (high) | Naturally occurring (low, ~1 muon/$cm^2$/minute) |
| Use collimator | Yes | No |
| Path type | Straight | Non-straight |
| Integral information | $\ln(\frac{I_0}{I})$ | $\ln(\frac{I_0}{I})$ or $\sigma_\theta^2$ |
| Modality | Transmission | Scattering or Transmission |
| Reconstruction methods | FBP, Iterative (ART, SIRT, SART) | FBP, Iterative (ART, SIRT, SART) |

## III. FRAMEWORK FOR MUON CT

In x-ray CT, let $I_0$ and I be the incident and outgoing beam intensity, respectively. The ratio $\ln(I_0/I)$ is used to reconstruct an object under investigation using FBP or ART as shown in Fig. 1 [17]:

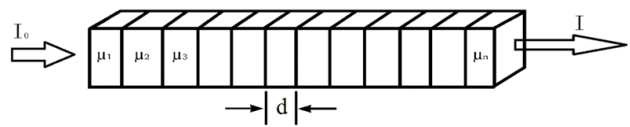

Fig. 1. Illustration of neutral beam crossing a discretized object

In Fig. 1, the attenuated intensity I can be described by:

$$I = I_0 e^{-d \sum_{i=1}^{n} \mu_i}, \qquad (1)$$

where d is a selected discretized length in cm and $\mu_i$ is the attenuation coefficient of the $i^{th}$ pixel in $cm^{-1}$. After rearrangement,

$$\ln(\frac{I_0}{I}) = d \sum_{i=1}^{n} \mu_i. \qquad (2)$$

The signal obtained from one projection or view is not enough to reconstruct an image. One typically rotates the radiation source and the detectors, while the object remains stationary, to obtain additional information from different angles.

In µCT, the incident source is naturally occurring cosmic ray muons. When muons traverse an object, many different scattering angles are registered that approximately follow a Gaussian distribution with zero mean value and variance given by [28]:

$$\sigma_\theta \cong \frac{15 MeV}{\beta c p} \sqrt{\frac{L}{L_{rad}}}, \quad (3)$$

where $p$ is the muon's momentum in MeV/c, $L$ is the length of the object, and $L_{rad}$ is the radiation length of the material. The concept of a muon traversing an object is shown in Fig. 2 [17]:

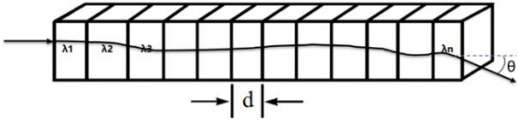

Fig. 2. Illustration of a muon traversing a discretized object. The magnitude of the scattering angle is exaggerated in the figure for illustration purposes.

The variance of scattering angle of a monoenergetic muon beam caused by the $i^{th}$ voxel can be written:

$$\sigma_{\theta_i}^2 = d\lambda_i, \quad (4)$$

where $\lambda_i$ is the scattering density of the $i^{th}$ pixel. The scattering density is defined as [8]:

$$\lambda(L_{rad}) \equiv \left(\frac{15}{p_0}\right)^2 \frac{1}{L_{rad}}, \quad (5)$$

where $p_0$ is the nominal momentum. Since MCS in individual voxels can be treated as independent, the variance of the scattering angle of a muon beam after traversing the entire object may be written as:

$$\sigma_\theta^2 = d \sum_{i=1}^{n} \lambda_i. \quad (6)$$

Note that Eqs.
(2) and
(6) have the same form; i.e., the right side of these two equations is a linear integration of a parameter over the particle's path. Thus, the scattering density λ may be treated similar to the attenuation coefficient μ used in the x-ray computed tomography image reconstruction process.

In order to obtain parallel or quasi parallel muon beams, the detectors are rotated around the object under interrogation. Let the position of $i^{th}$ muon be $(x_{1i}, y_{1i}, z_{1i})$, $(x_{2i}, y_{2i}, z_{2i})$, $(x_{3i}, y_{3i}, z_{3i})$, and $(x_{4i}, y_{4i}, z_{4i})$ on four detectors separately, two before the object and two after. The incident azimuth angle of the $i^{th}$ muon $\varphi_i$ is

$$\varphi_i = \arctan\left(\frac{y_{2i} - y_{1i}}{x_{2i} - x_{1i}}\right). \quad (7)$$

The scattering angles $\theta_i$ were calculated using:

$$\theta_{iy} = atan\left(\frac{y_{4i} - y_{3i}}{z_{4i} - z_{3i}}\right) - atan\left(\frac{y_{2i} - y_{1i}}{z_{2i} - z_{1i}}\right) \quad (8)$$

$$\theta_{iy} = atan\left(\frac{y_{4i} - y_{3i}}{z_{4i} - z_{3i}}\right) - atan\left(\frac{y_{2i} - y_{1i}}{z_{2i} - z_{1i}}\right)$$

$$\theta_i = \sqrt{\frac{\theta_{ix}^2 + \theta_{iy}^2}{2}}$$

Using each muon's momentum to correct for the influence of polyenergetic muons, in this work the nominal momentum $p_0$ is chosen to be 3 GeV/c. The quantity $p_i$ is the initial momentum, and it is assumed that there are no energy losses during the process of crossing objects. If momentum information is not available, this step will be simply omitted:

$$\theta_i' = \frac{p_i}{p_0} \theta_i, \quad (9)$$

Due to the uniformity along the Z axis of our imaging object, after incorporating the path length into correction for the influence of different path length along vertical direction, the normalized scattering angle of a muon becomes

$$\theta_i'' = \sqrt{\frac{L_{ih}}{L_i}} * \theta_i', \quad (10)$$

where $L_i$ is the distance between point $(x_{2i}, y_{2i}, z_{2i})$ and point $(x_{3i}, y_{3i}, z_{3i})$, and $L_{ih}$ is the horizontal projection of $L_i$. Finally, the registered incident muon spectrum is divided into one-degree-wide azimuthal bins according to their incident azimuth angles φ, re-sorting the incident muons into 180 quasi-parallel groups and projecting to a horizontal plane.

## IV. MUON TRACING AND SCATTERING ANGLE PROJECTION ALGORITHM

Three different muon tracing algorithms are proposed and investigated in this section. These are: (1) use of a straight path along a muon's incident trajectory, (2) use of a straight path along a muon's incident direction crossing its PoCA point, and (3) use of a muon's POCA trajectory. For each tracing algorithms, two different scattering angle projections were used: (a) each scattering angle is used only once and stored directly into the corresponding detector bins, or (b) each scattering angle is back projected into the image space first to calculate the variance of the scattering angle in each pixel, then the summation of the variances is forward projected to the corresponding detector bins. The three muon tracing algorithms and two scattering angle projections result in six combinations: **1a**, **1b**, **2a**, **2b**, **3a,** and **3b.** These methods were used to generate projection information and corresponding system matrices to investigate how muon ray tracing methods and scattering angle projection methods affect muon CT reconstruction image quality and detection capability.

### A. Tracing algorithm 1: Use of a straight path along the incident muon trajectory (straight path tracing)

This method assumes that a muon experiences no scattering events or that scattering is negligible. This results in a straight path crossing the object along the muon incident trajectory, and completely ignoring the exit position. The scattering angle is back-projected in two different ways: method **1a** is used to directly store the scattering angles for each muon from the same quasi-parallel beam subset into the corresponding detector bins hit by its path, then the variance is calculated from the scattering angles in each bin and used as projection information $P$. Method **1b** is used to back project each scattering angle into the pixels crossed by this straight path, for all muons in the same quasi-parallel beam subset, and then to calculate the variance of the scattering angle in each pixel, and, finally, it takes the summation of the variances along this path and stores it into the corresponding detector bin as projection information $P$. Either filtered back projection (FBP) or a simultaneous algebraic reconstruction technique (SART) can be used to reconstruct an image using $P$. For FBP, one simply applies a high pass filter to the projection information $P$ stored in the detector bins before back projecting it into the space domain. For SART, the average path length of the muons in the same quasi-parallel beam subset in each pixel is used to build a system matrix $W$ as shown in Fig. 3.

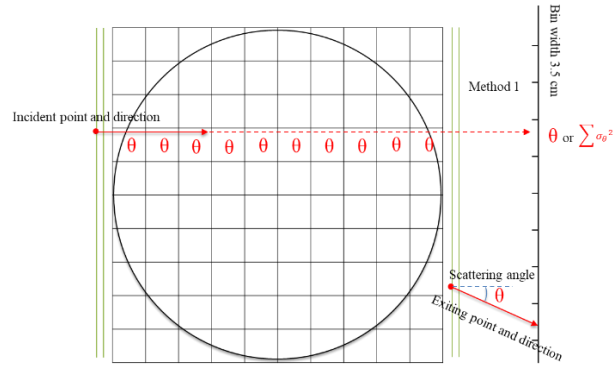

Fig. 3. Top-down view illustration of tracing algorithm **1** (straight path tracing). Approach **1a** projects the scattering angle within a defined volume (shown as a large square, the black circle indicates dry storage cask) along a straight line to a segmented detector (shown at far right). In approach **1b**, $\theta_i$ is back projected into the pixels crossed by this straight path indicated by the red dashed line. See the text for details

### B. Tracing algorithm 2: Use of a straight path along the muon incident direction that crosses the PoCA point (PoCA point tracing)

This algorithm assumes that a muon experiences a single Coulomb scattering event within a defined volume. The scattering angle is taken to occur at the closest distance between the incident and exiting trajectories. This point is also known as the PoCA point. Instead of completely ignoring the exit position, tracing algorithm **2** makes a compromise between a muon's incident and exiting positions by assuming the muon crossed the object along incident direction crossing the PoCA point, as shown in Fig. 4. The rest of the steps are the same with tracing algorithm **1**.

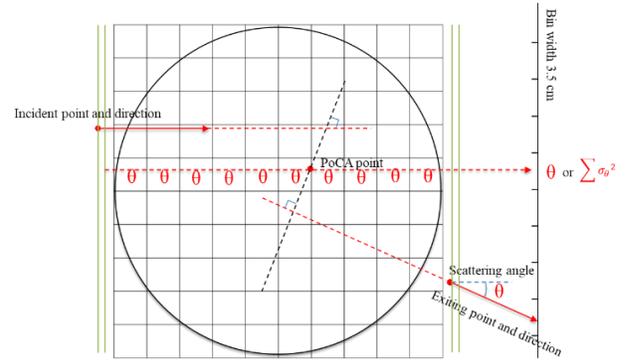

Fig. 4. Top-down view illustration of tracing algorithm **2** (PoCA point tracing). Approach **2a** projects the scattering angle within a defined volume (shown as the large square, the black circle indicates dry storage cask) along a straight line to a segmented detector (shown at right). In approach **2b**, $\theta_i$ is back projected into the pixels crossed by this straight path indicated by the red dashed line. See the text for details

### C. Tracing algorithm 3: Use of PoCA trajectory (PoCA trajectory tracing)

Since it is a charged particle, a muon experiences multiple Coulomb scattering while it traverses objects, causing it to deviate from a straight path. Thus, a curve path may better approach muons trajectory crossing objects than a simple straight line. This tracing algorithm assumes that a muon travels along the so-called PoCA trajectory within our defined volume. The PoCA trajectory consists of two segments: (1) the segment connecting the point of muon incidence to the PoCA point and (2) the segment connecting the PoCA point and the point at which the muon exits said volume, as descried in Fig. 5. The rest of the steps are the same as those described in tracing algorithm **1**.

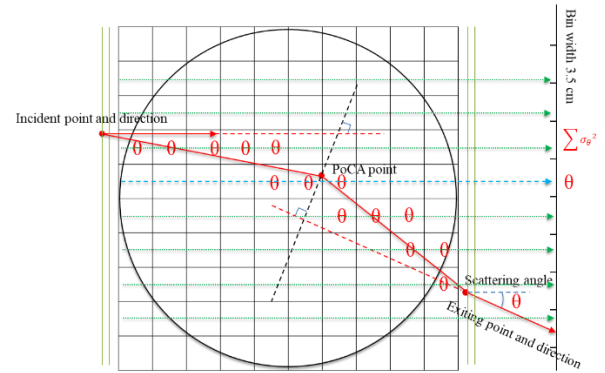

Fig. 5. Top-down view illustration of tracing algorithm **3** (PoCA trajectory tracing). According to apporach **3b**, the scattering angle is projected back into the pixels crossed by the PoCA trajectory for all muons in a quasi-parallel beam subset, the variance in each pixel is calculated, and the sum of the variances of scattering angle along its incident horizontal direction indicated by green line is stored in the corresponding detector bins. Approach **3a** projects the scattering angle along the blue dashed line to a segmented detector (shown at far right), and the system matrix is calculated with the PoCA path indicated by the red segments

## V. Image Reconstruction

In this work, the reconstruction was implemented simultaneous algebraic reconstruction technique (SART) [29]. Conventional FBP was also implemented for comparison purposes. Each projection may be analytically expressed as an integration of the scattering density along the muon path:

$$P_\theta(R, \theta) = \iint f(x, y) dx dy, \quad (11)$$

where $f(x, y)$ is the scattering density at position $(x, y)$. Similarly, if we discretize the reconstruction volume, this could be expressed in the fashion of a matrix as

$$WX = P, \quad (12)$$

where $W$ is the system matrix containing the average path length, $X$ is the scattering density map of the object to be reconstructed, and $P$ is the projection information. SART can be used to solve Eq. (12), which is [29]:

$$X_j^{(K+1)} = X_j^{(K)} + \frac{\lambda \sum_i [w_{ij} \frac{p_i - \vec{w}_i^T X^{(K)}}{\sum_{j=1}^N w_{ij}}]}{\sum_i w_{ij}} \quad (13)$$

A total of $N=90$ angular views were used. The object to be reconstructed was discretized into 100×100 pixels, and the scattering density was expressed as a $100^2 \times 1$ vector $X$. The iterative reconstruction process was stopped when the maximum iteration number was reached or when the difference in successive iterations was below a threshold. In this paper, relaxation $\lambda = 0.45$ and a max of 100 iterations were chosen. A pseudocode is shown in Fig. 6.

## VI. Validation Against Experimental Results

Prior to using muon simulated data for μCT reconstruction of shielded objects, a GEANT4 model was developed and validated against a physical experiment recently done by Los Alamos National Laboratory (LANL) [14]. LANL used a muon detector system to take measurements of a partially loaded MC-10 steel dry storage cask located at the Idaho National Laboratory. The cask was filled with 18 out of 24 PWR fuel assemblies, as shown in Fig. 7 on the left. Each fuel assembly is about 21.4 cm on each side. Muon tracking detectors were placed on opposite sides with a relative elevated difference of 1.2 m. A total of 9 measurement configurations were performed by placing the detectors at different horizontal positions. Each configuration collected between $4 \times 10^4$ to $9 \times 10^4$ muons.

For validation purposes, the exact geometry of the MC-10 dry storage cask and the detector configuration were simulated in GEANT4 in an effort to reproduce as accurately as possible the actual experimental configuration. The simulated configuration is shown in Fig. 7 on the right. The muon detectors were shifted 6 times in total following the experimental procedure described in [14]. Radiation emitted from the cask was not simulated, and the detectors were modelled to have 100% efficiency.

---

1) Collect incident and exiting muon position $(x_{1i}, y_{1i}, z_{1i})$, $(x_{2i}, y_{2i}, z_{2i})$, $(x_{3i}, y_{3i}, z_{3i})$, and $(x_{4i}, y_{4i}, z_{4i})$ for M views
2) Calculate scattering angle $\theta$, PoCA point and azimuth angle for each muon.
3) Correct scattering angle $\theta_i$ with path length $L_i$ and muon momentum $p_i$ (if available).
4) Re-sort muons into quasi-parallel subsets according to their azimuth angle $\varphi$.
5) Discretize the reconstruction volume into N×N pixels.
6) For each muon quasi-parallel subset,
   if use projection method **a**, do:
   a) Calculate the path length in pixels crossed by each muon in this subset using of the three trajectory models. At the end of the last muon in this group, take the average path length in each pixel and store them in system matrix $W$.
   b) Project scattering angle $\theta''$ into corresponding detector bins along it incident direction. At the end, calculate the variance $\sigma_\theta^2$ of scattering angle in each bin and store them in $P$.
   if use projection method **b**, do:
   c) Similar to 6a.
   d) Back project the scattering angle $\theta''$ of each muon in this subset into pixels crossed by its trajectory. At the end of last muon in this group calculate the variance of scattering angle in each pixel
   e) take summation of variance of scattering in pixels along incident horizontal direction of beam and store it in $P$
7) Solve equation $WX = P$ using SART or SIRT with or without regularization.

Fig. 6. A pseudocode for muon computed tomography

A simple PoCA method was used to generate a muon scattering angle map of the dry storage cask. The mean scattering angle was then calculated, and it was compared to the reported experimental results. Fig. 8 shows a comparison of experimental and simulated values of the average scattering angle in a slice horizontally crossing the center of the cask. The results show that our simulation is in good agreement with the experimental measurements, except in the region from -20 cm to 40 cm where a maximum discrepancy of 3 mrad is observed. The cask wall and columns no. 2, 3, 4, and 5 are in good agreement with the experimental measurements. One would expect column 1 to have lower experimental values than column 2 given that it holds no assemblies. This effect is captured by the simulation but not by the experiment. A possible reason that this was not captured in the experimental measurements is a reported detector motion due to strong winds

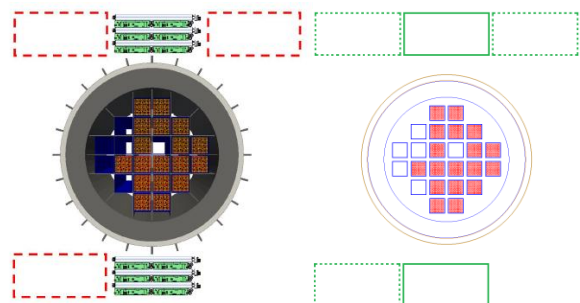

Fig. 7. MC-10 cask configuration used in LANL experiment [15] (left) and GEANT4 simulated cask (right). The simulated detector positions are shown in green, the fuel assemblies in red, and the empty locations in blue.

at the time of the experiment.

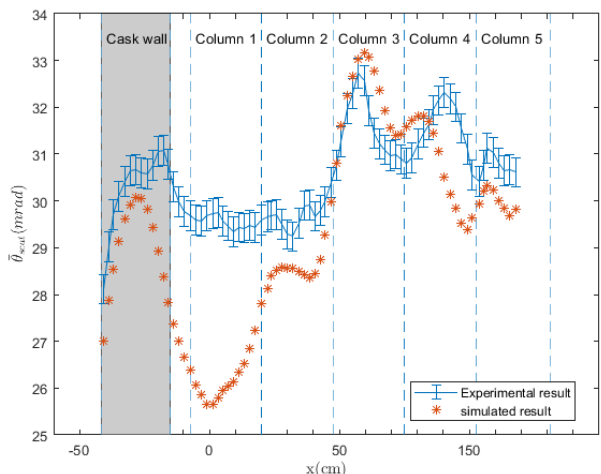

Fig. 8. Experimentally measured [15] (blue line) and GEANT4 simulated (orange circles) average scattering angles for muons crossing a MC-10 dry storage cask.

## VII. NUMERICAL RESULTS AND DISCUSSION

### A. Test model configuration

To illustrate and evaluate the proposed muon tracing algorithms, we simulated a shielded VSC-24 dry cask storing commercial spent nuclear fuel assemblies. The VSC-24 is described in detail in reference [25]. This cask contains fuel assemblies in a thin steel canister that is shielded externally by a thick concrete overpack. This configuration is notably different than the MC-10 steel cask measured by LANL; a VSC-24 type concrete cask is widely used for storing spent nuclear fuel and was selected to better represent the existing dry storage cask population. The VSC-24 simulated dry cask geometry is shown in Fig. 9. The dry cask was fully loaded with one fuel assembly missing from row 3. Two pairs of identical planar detectors with dimensions 350 cm×150 cm, vertically offset by 100 cm, and positioned along the sides of the dry cask, were simulated using GEANT4. For each pair, the separation between each planar detector was 10 cm. The zenith angle was ~50°, yielding a muon flux of ~20,000 muons/min.

The simulated detectors [30] were modeled as planes with perfect efficiency, spatial and energy resolution. The muon event generator described in [31] was used to simulate the muon flux at sea level. In our simulated implementation, the cask containing the spent fuel assemblies was fixed, and the detectors rotated around it. The detectors were rotated at 2° increments to collect data from multiple views.

### B. Results

A quantity of $1.8 \times 10^7$ muons were generated and used for the reconstruction of the simulated dry cask, which is equivalent to ~48 hours of exposure. Tracing algorithms **1a, 1b, 2a, 2b, 3a,** and **3b** were used for muon tracing and data processing. The image reconstructions were performed using SART, but FBP was also used with straight path tracing algorithms **1a** and **1b** for comparison purposes. The results reconstructed with FBP, using **1a** or **1b** with and without muon momentum information are shown in Figs 10 and 11. FBP performs reasonably well with the missing assembly visible in all cases. The concrete overpack and steel canister can also be identified. A significant deterioration in image resolution is observed for the scenario where muon momentum is not available.

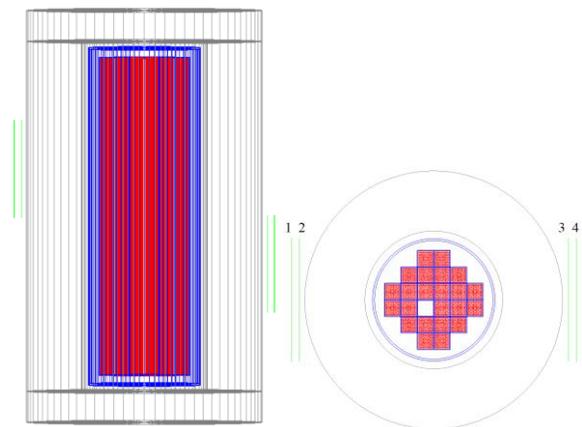

Fig. 9. Side (left) and top-down (right) illustrations of the simulated cask and muon detectors built in Geant4. An assembly has been removed from column 3. Muon detectors are shown in green, fuel assemblies in red. Blue and grey are used to represent the steel canister and concrete overpack, respectively.

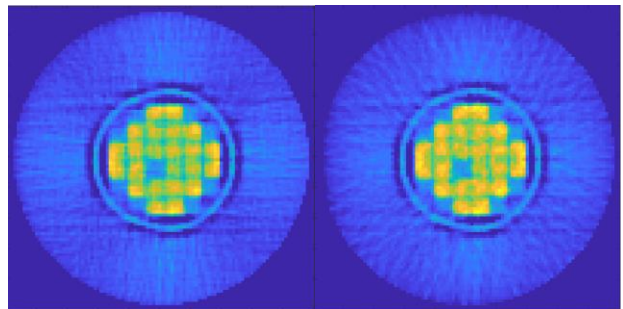

Fig. 10. FBP reconstruction of a dry cask (refer to Fig. 9) with perfect momentum measurement. Results are shown using straight path tracing **1a** on the left and **1b** on the right.

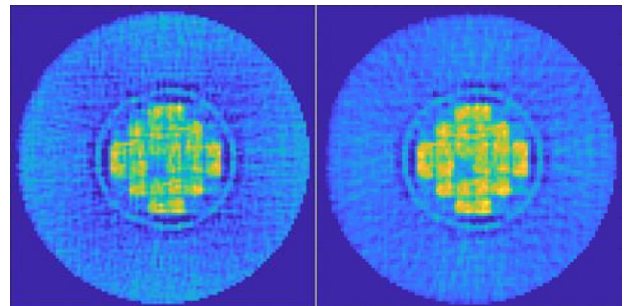

Fig. 11. FBP reconstruction of a dry cask (refer to Fig. 9) without momentum measurement. Results are shown using straight path tracing **1a** on the left and **1b** on the right.

The results reconstructed with SART, using tracing algorithms **1, 2** and **3** with and without muon momentum information are shown in Fig. 12 and Fig. 13. Similarly to FBP, the location of the missing fuel assembly is easily identifiable. However, the image quality is improved when compared to FBP, even without muon momentum information. In addition, it appears that all tracing algorithms have improved image quality when compared to FBP with the algorithm **3b** to have the best image quality.

Signal to noise ratio (SNR), contrast to noise ratio (CNR) and detection power (DP) were used to quantitatively assess how the muon tracing and scattering angle projection algorithms would be expected to affect the reconstructed image quality. SNR, CNR, and DP were calculated as [32]:

$$\text{SNR} = \frac{\text{mean}(8 \text{ assemblies surrounding missing one})}{\text{std}(8 \text{ assemblies surrounding missing one})}$$

$$\text{CNR} = \frac{\text{mean}(8 \text{ assemblies}) - \text{mean}(\text{missing one})}{\max(\text{std}(8 \text{ assemblies}), \text{std}(\text{missing one}))}$$

$$\text{DP} = \text{SNR} * \text{CNR}$$

where std denotes standard deviation. There are two different regions of interest in the model: (i) the empty slot and (ii) the surrounding 8 spent nuclear fuel assemblies. The SNR is used to quantify signal strength and also how uniform the estimated scattering density is within the assemblies. CNR is used to quantify how different these two regions are. A large CNR means the reconstruction method is more capable to differentiate between the two regions. Since signal strength, uniformity and contrast are all important in detection of anomalies, the multiplication of SNR and CNR is defined to be detection power (DP) of the reconstruction method. The SNR, CNR and DP of the reconstructed images shown in Fig. 10 to Fig. 13 are given in Table II.

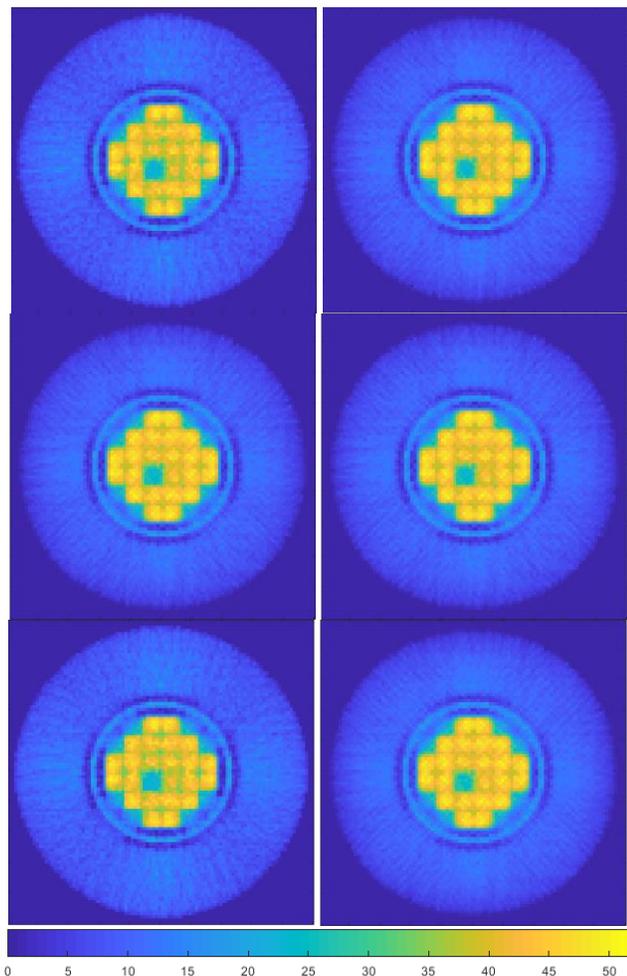

Fig. 12. SART reconstruction of a dry cask (refer to Fig. 9) with perfect momentum measurement. Results are shown using tracing algorithm **1a** (top-left), **2a** (middle-left), **3a** (bottom-left), **1b** (top-right), **2b** (middle-right), and **3b** (bottom-right).

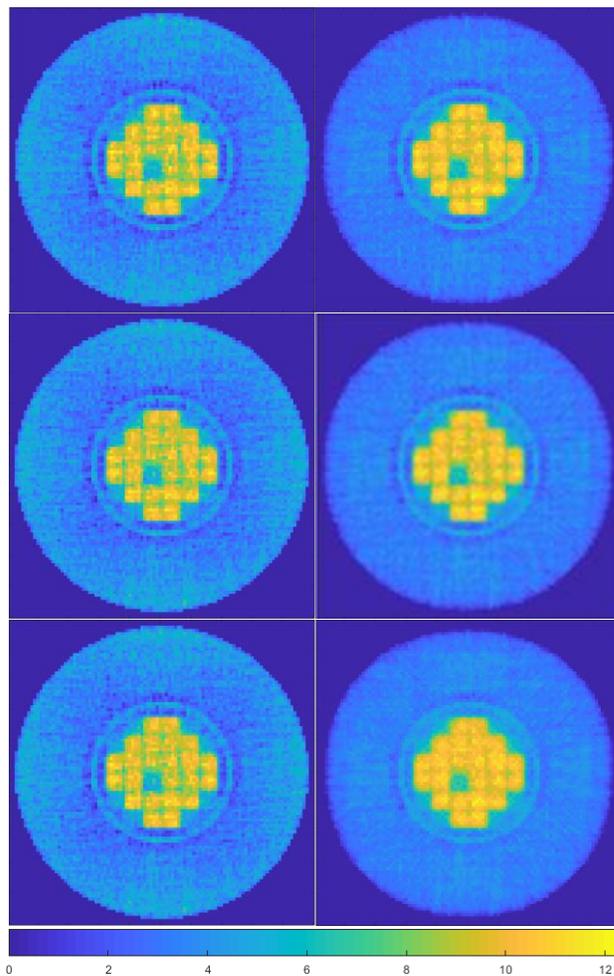

Fig. 13. SART reconstruction of a dry cask (refer to Fig. 9) without momentum measurement. Results are shown using tracing algorithm **1a** (top-left), **2a** (middle-left), **3a** (bottom-left), **1b** (top-right), **2b** (middle-right), and **3b** (bottom-right).

A comparison of expected detection power for methods using path **1**, **2** or **3** and projection method **a** or **b** is shown in Fig. 14. A value of 0 on the path type axis represents the method using FBP and is used as baseline. The rest (path type 1, 2, and 3) are based on SART (refer to Table II).

TABLE II
EXPECTED IMAGE CHARACTERISTICS FOR THE DESCRIBED METHODS

| Tracing algorithm | With momentum | | | Without momentum | | |
|---|---|---|---|---|---|---|
| | SNR | CNR | DP | SNR | CNR | DP |
| | FBP | | | | | |
| 1a | 6.30 | 2.73 | 17.20 | 5.13 | 1.98 | 10.17 |
| 1b | 7.00 | 2.60 | 18.25 | 6.36 | 1.94 | 12.35 |
| | SART | | | | | |
| 1a | 11.94 | 5.06 | 60.46 | 9.27 | 3.82 | 35.39 |
| 2a | 12.22 | 5.29 | 64.73 | 9.43 | 3.89 | 36.68 |
| 3a | 12.19 | 5.28 | 64.32 | 9.41 | 3.88 | 36.50 |
| 1b | 14.06 | 5.68 | 80.04 | 12.96 | 4.94 | 64.10 |
| 2b | 13.97 | 5.68 | 79.22 | 12.40 | 4.64 | 57.59 |
| 3b | 14.54 | 5.60 | 81.45 | 15.64 | 4.96 | 77.61 |

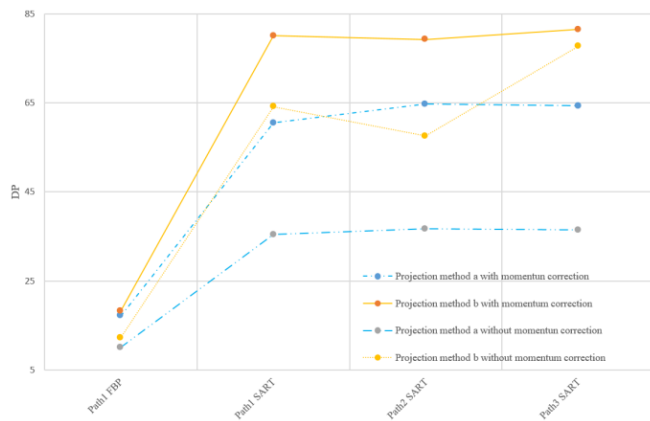

Fig. 14. A comparison of the expected results for the different described methods with and without momentum information. See text for details.

In both Table II and Fig. 14, it can be observed that the DP for the straight path along the muon incident trajectory (path type 1), the straight path along the muon incident direction crossing the PoCA point (path type 2), and the PoCA trajectory (path type 3), the reconstructed image quality (SNR and CNR) is quite similar and higher than FBP (path type 0) in all cases. When muon momentum is not available, the projection method **3b** is 23.4%, 3.5%, and 27.6% better than FBP in terms of CNR, SNR or DP, respectively. Further, scattering angle projection method **b** is higher in performance by 45.6%, 25.7%, and 83.6% when compared to projection method **a** in terms of CNR, SNR, or DP. By comparing the cases with and without momentum in Fig. 14, on average when using projection method **a** would have a 42.7% detection power loss when muon momentum information is absent; however, on average the detection power when using projection method **b** only decreases by 17.1%. Thus, by back projecting a muon's scattering angles to the pixels crossed by its trajectory and calculating the variance of the scattering angle in each pixel and then taking the summation of the variance along the incident direction as the projection information could significantly improve the expected image quality and detection power while reducing the reliance on muon momentum information. Due to the difficulty in precisely measuring muon momentum in an experiment, it is expected that the PoCA trajectory (path type 3) in combination with projection method **b** and SART reconstruction would achieve the best performance among the methods discussed.

### C. Detection Limit

In order to investigate the expected detection limit [13] of our µCT algorithms in scenarios that are relevant to international safeguards, a dry storage cask with partial defects in assemblies was simulated. Without loss of generality, we investigated a geometry where one half and a quarter of selected fuel assemblies were missing at different locations. In Fig. 15 (see top left), one can see the location of removed quarter and half assemblies. One quarter assembly was removed at the center, two quarter fuel assemblies were removed at the edge, and one-half assembly was removed at the edge. A quantity of $1.5 \times 10^7$ muons, equivalent to ~39.5 hours of exposure, was used to reconstruct the dry cask with algorithm **3b**. In reconstructed images, with or without momentum information, the location of all missing fuel assemblies, quarter and half, appears to be visible.

To analyze this observation quantitatively, two ways to detect missing fuel were investigated. The first way exploited the geometric symmetry of the 24 spent nuclear fuel assemblies. Six slices crossing the reconstructed image using algorithm **3b** but without momentum are shown in Fig. 15. A comparison between slice 1 and slice 3 was used to detect a partial defect at the edge. Similarly, slice 2 reveals the missing fuel at the center of the cask, and the comparison between slice 4 and slice 5 reveals a partial defect on the rim between two fuel assemblies. The second way was to calculate the difference in scattering density between an empty quarter slot and the rest of the same fuel assembly. It turned out that a quarterly fuel missing at the upper left edge was the most difficult to detect among these locations (see Fig. 15). The estimated scattering densities in this missing quarter fuel slot and the remaining three quarters of the fuel were $7.3 \pm 0.8$ and $10.2 \pm 0.5$ (arb. units), respectively; these values are separated by a difference of 5.8 σ. Of course, actual measurement conditions, radiation background, and measurement errors would reduce this difference. However, one may expect that the limiting case of a quarter fuel assembly missing to be detected given sufficient measurement time to allow enough muons to be measured.

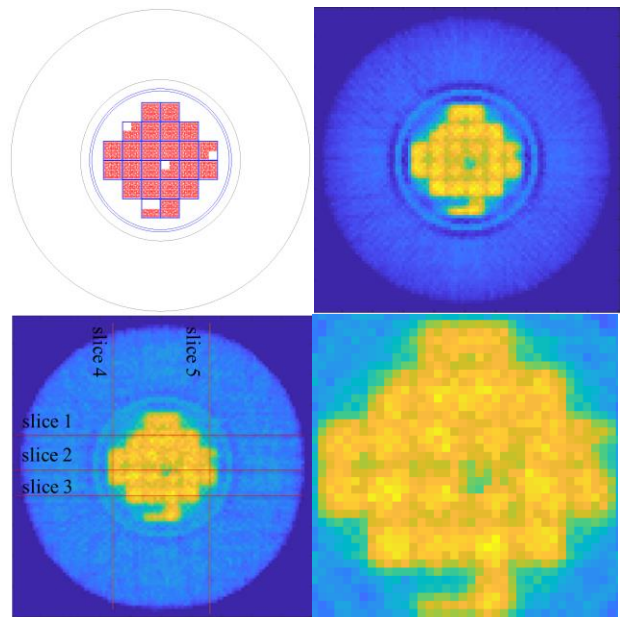

Fig. 15. Top-down (top-left) view of the cask with a half assembly and 3 quarter assemblies missing. The reconstructed image using method **3b** is shown with perfect momentum information (top-right), without momentum information (bottom-left), and in a zoomed-in view without momentum information (bottom-right).

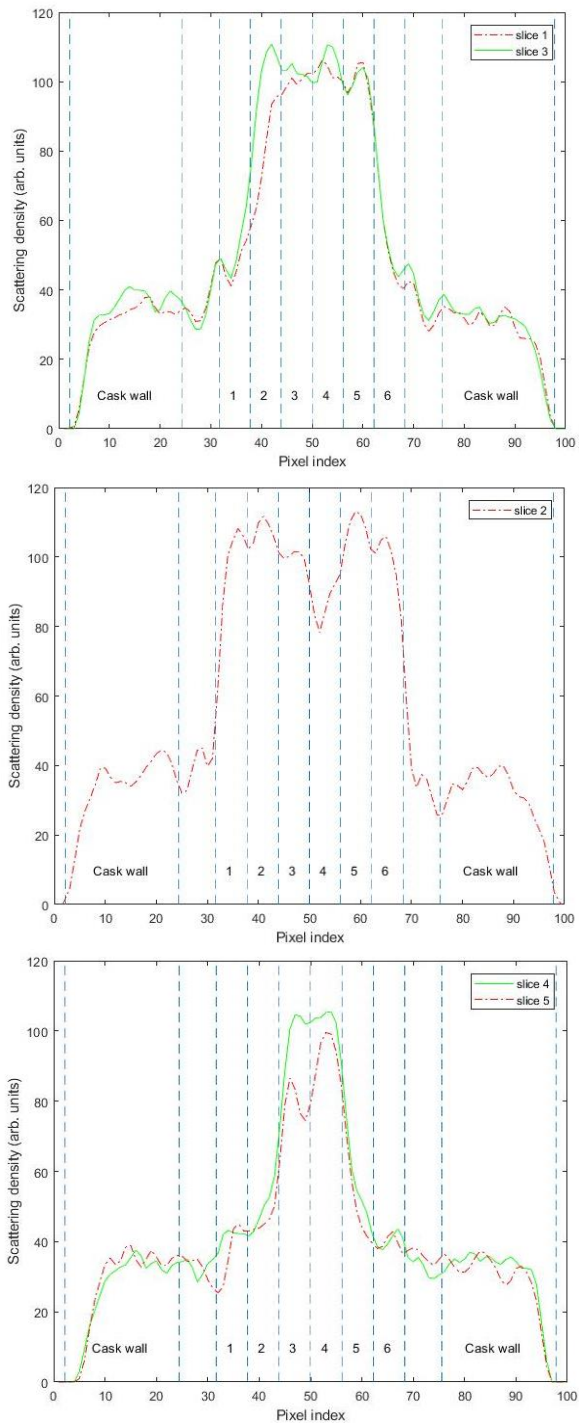

Fig. 16. A comparison of the simulated slices of scattering density, showing slice 1 and 2 at the top, slice 2 in the middle and a comparison of slice 4 and 5 at the bottom. The slices are illustrated in Fig. 15. See text for details.

*D. Detector Size*

The use of large area, position-sensitive planar or ring-shaped detectors able to cover the whole cask can generate complete information about the contents of the cask wall and the spent nuclear fuel. However, it may not be economically practical to build such large area detectors with a large number of readout channels [14, 15, 31]. Smaller detectors whose width is equal or slightly less than the diameter of steel canister should still yield valuable information about the fuel location within the cask. Due to the central symmetry of the cask wall and its smaller scattering density compared to spent nuclear fuel in the middle, complete sinogram information of the whole cask is not necessary.

To support this assertion, planar muon detectors having an area of $1.6 \times 1.2$ m$^2$ were simulated to register muons crossing the aforementioned VSC-24 dry cask. The new detector size is shown on the left in Fig. 17. Only the data registered by the muon tracks passing through the $1.6 \times 1.2$ m$^2$ detector area was used to reconstruct the spent nuclear fuel using method **1a**. The expected result is shown in Fig. 17. Strong strip artifacts due to the steel canister are expected when using small size detectors. After removing the overpack from the image, the spent nuclear fuel assemblies are highlighted and clearly visible, as shown at on the right in Fig. 17. The expected SNR, CNR, and DP are 8.73, 4.87, and 42.53, respectively. For the same equivalent acquisition time (47.4 hours of exposure, $6.6 \times 10^6$ muons), these values are smaller than the corresponding values reconstructed with the simulated large area detectors: 11.94, 5.06, and 60.46, respectively. Even so, the reconstructed signal in the empty slot is still expected to be significantly separated by the surround slots by 4.87σ, (large area detectors: 5.06σ). Thus, smaller detectors are expected to be able detect a single missing fuel assembly with good confidence.

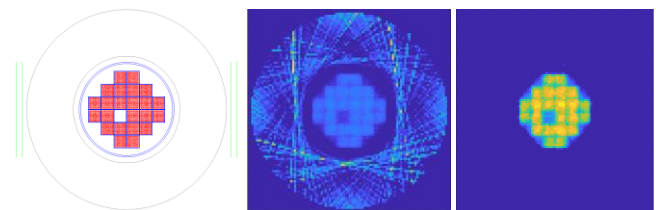

Fig. 17. Top-down illustration of the cask and detectors built in Geant4 (top-left), original reconstructed image (top-right) with method 1**a**, and the reconstructed image of the spent nuclear fuel after removing the image of cask (bottom-left).

## VIII. CONCLUSION

In this paper, three different muon tracing methods (use of muon incident trajectory, straight path along its incident direction crossing its PoCA point, or PoCA trajectory) along with two scattering angle projection methods (scattering angles projection to the muon detector bins or projection of the summation of scattering density to the corresponding detector bins), combined together as 6 different methods, were investigated. These methods are generally applicable for use in μCT, and they were applied to reconstruct a VSC-24 dry storage cask. A GEANT4 simulation workspace was developed and compared against the only available experimental data from a MC-10 dry storage cask. The results showed reasonable agreement with experimental measurements, providing confidence in our simulations. For the VSC-24 model, FBP and ART-based reconstruction methods were used to reconstruct the projected information stored in the detector bins. Algebraic reconstruction techniques were shown to be more useful in this application since muon trajectories are not generally straight. When muon momentum information is not available, use of the PoCA trajectory along with scattering angle projection method **b** is expected to greatly outperform the conventional use of straight path trajectories (see Fig. 14). Additionally, projection method **b** is expected to improve anomaly detection power and reduce the reliance on muon momentum information. Further, a simulated VSC-24 dry cask with portions of assemblies missing was used to analyze the expected detection limit. Method **3b** is expected to be able to detect an aggregate quarter of a missing assembly at any location in the cask, even without any momentum information. Finally, it is expected that detectors having width equal to the diameter of the steel canister inside the concrete overpack can reconstruct the spent nuclear fuel assemblies and identify missing assemblies.

ACKNOWLEDGMENT

This research was performed using funding received from the DOE Office of Nuclear Energy's Nuclear Energy University Programs under contract DE-NE0008292. This research was also partially sponsored by the Laboratory Directed Research and Development Program of Oak Ridge National Laboratory, managed by UT-Battelle, LLC, for the US Department of Energy. We would like to thank Dr. J. M. Durham for useful discussions regarding our GEANT4 model validation against LANL's physical experiment on a MC-10 dry storage cask.

REFERENCES

[1] T. K. Gaisser et al., "Cosmic rays and particle physics," Cambridge University Press, 2016.
[2] K. Hagiwara, "Review of particle physics," Phys. Rev. D, vol. 66, no. 1, p. 010001, 2002.
[3] E. Guardincerri et al., "Imaging the inside of thick structures using cosmic rays," AIP Advances vol. 6, 015213, 2016.
[4] E. Guardincerri, et al., "3D Cosmic Ray Muon Tomography from an Underground Tunnel," Pure and Applied Geophysics vol. 174.5, 2133-2141, 2017.
[5] K. Borozdin, et al., "Cosmic ray radiography of the damaged cores of the Fukushima reactors," *Physical review letters* vol. 109.15, 152501, 2012.
[6] P. Baesso et al., "Toward a RPC Based Muon Tomography System for Cargo Containers," J. Instrum., vol. 9, C10041, 2014.
[7] C. L. Morris et al., "Obtaining material identification with cosmic ray radiography," AIP Adv., vol. 2, no. 4, p. 042128, 2012.
[8] L. J. Schultz, "Cosmic ray muon radiography," Ph.D. dissertation, Dept. Elect. Comput. Eng., Portland State Univ., Portland, OR, USA, 2003.
[9] E. Aström et al., "Precision measurements of linear scattering density using muon tomography," J. Instrum., vol. 11, p. P07010, Jul. 2016.
[10] S. Chatzidakis, "Cosmic Ray Muons for Spent Nuclear Fuel Monitoring," Ph.D. Dissertation, Dept. Nucl. Eng., Purdue University, IN, USA, 2016.
[11] S. Chatzidakis et al., "A Bayesian approach to monitoring spent fuel using cosmic ray muons," Trans. Amer. Nucl. Soc., vol. 111, pp. 369–370, 2014.
[12] S. Chatzidakis et al., "Monte-Carlo simulations of cosmic ray muons for dry cask monitoring," Trans. Amer. Nucl. Soc., vol. 112, pp. 534–536, 2015.
[13] S. Chatzidakis, C. K. Choi, and L. H. Tsoukalas, "Interaction of cosmic ray muons with spent nuclear fuel dry casks and determination of lower detection limit," Nucl. Instrum. Methods, Phys. Res. A, vol. 828, pp. 37–45, Aug. 2016.
[14] J. M. Durham et al., "Cosmic ray muon imaging of spent nuclear fuel in dry storage casks," J. Nucl. Mater. Manage., vol. 44, p. 3, 2016.
[15] J. M. Durham et al., "Verification of spent nuclear fuel in sealed dry storage casks via measurements of cosmic ray muon scattering," Phys. Rev. Applied, vol. 9, p. 044013, 2018.
[16] S. Chatzidakis et al., "Analysis of Spent Nuclear Fuel Imaging Using Multiple Coulomb Scattering of Cosmic Muons," IEEE Trans. Nuc. Sci., vol. 63, p. 2866, 2016.
[17] Z. Liu et al., "Detection of Missing Assemblies and Estimation of the Scattering Densities in a VSC-24 Dry Storage Cask with Cosmic-Ray-Muon-Based Computed Tomography," J. Nucl. Mater. Manage., vol 45, p. 12, 2017.
[18] D. Poulson et al., "Cosmic ray muon computed tomography of spent nuclear fuel in dry storage casks," Nuclear Instruments and Methods in Physics Research Section A: Accelerators, Spectrometers, Detectors and Associated Equipment, vol. 842, pp. 48-53, 2017.
[19] Z. Liu et al., "Characteristics of Muon Computed Tomography of Used Fuel Casks Using Algebraic Reconstruction," IEEE Nucl. Sc. Symp. Conf. Record, 2017.
[20] L. J. Schultz et al., "Image reconstruction and material Z discrimination via cosmic ray muon radiography," Nucl. Instrum. Methods, Phys. Res. A, vol. 519, no. 3, pp. 687–694, 2004.
[21] G. Charpak et al., "Applications of 3-D nuclear scattering radiography in technology and medicine," Proc. SPIE, vol. 0312, pp. 156–163, Aug. 1983.
[22] H. A. Bethe, "Moliere's Theory of Multiple Scattering," Physical Review, vol. 89, p. 1256, 1953.
[23] V. L. Highland, "Some Practical Remarks on Multiple Scattering," Nucl. Inst. Meth., vol. 129, pp. 497–199, 1975.
[24] S. Chatzidakis et al., "A generalized muon trajectory estimation algorithm with energy loss for application to muon tomography," Journal of Applied Physics vol. 123, p. 124903, 2018.
[25] "Industry spent fuel handbook," Electr. Power Res. Inst., Palo Alto, CA, USA, Tech. Rep. 1021048, 2010.
[26] J. A. Sorenson and M. E. Phelps, Physics in nuclear medicine. New York: Grune & Stratton, 1987.
[27] D. W. Anderson, Absorption of ionizing radiation. University Park Press, 1984.
[28] A. A. M. Mustafa and D. F. Jackson, "Small-angle Multiple Scattering and Spatial Resolution in Charged

Particle Tomography," Phys. Med. Biol., vol. 26, pp. 461–472, 1981.
[29] A. C. Kak and M. Slaney, Principles of computerized tomographic imaging, IEEE press, 1988.
[30] V. Anghel et al., "A plastic scintillator-based muon tomography system with an integrated muon spectrometer," Nucl. Instrum. Methods, Phys. Res. A, vol. 798, pp. 12–23, Oct. 2015.
[31] S. Chatzidakis, S. Chrysikopoulou, and L. H. Tsoukalas, "Developing a cosmic ray muon sampling capability for muon tomography and monitoring applications," Nucl. Instrum. Methods, Phys. Res. A, vol. 804, pp. 33–42, Dec. 2015.
[32] J. L. Prince, and M. L. Jonathan, Medical imaging signals and systems. Upper Saddle River: Pearson Prentice Hall, 2006.